\documentclass[10pt]{ismd08}
\usepackage{graphicx}
\usepackage{amssymb}
\usepackage{cite,./mcite}

\newcommand{\bl}[1]{\begin{equation}\label{#1}}
\newcommand{\be}{\begin{equation}}
\newcommand{\ee}{\end{equation}}
\newcommand{\bea}{\begin{eqnarray}}
\newcommand{\eea}{\end{eqnarray}}

\newcommand{\rec}[1]{\frac{1}{#1}}
\newcommand{\td}[2]{\frac{\mathrm{d}{#1}}{\mathrm{d}{#2}}}
\newcommand{\z}[1]{\left({#1}\right)}

\newcommand{\kz}[1]{\left\{{#1}\right\}}
\renewcommand{\v}[1]{\mathbf{#1}}
\renewcommand{\c}[1]{\mathcal{#1}}

\renewcommand{\r}[1]{(\ref{#1})}
\begin{document}

\title{QCD EoS, initial conditions and final state from relativistic hydrodynamics in heavy-ion collisions}
\author{M\'arton Nagy$^1$\protect\footnote{\ \ speaker}}
\institute{$^1$MTA KFKI Research Institute for Particle and Nuclear Physics, Budapest, Hungary}
\maketitle

\begin{abstract}
Some recent developments in exact results in relativistic hydrodynamics is reviewed.
We discuss phenomenological applications in high-energy collisions and theoretical
features of the solutions. We compare the method of numerical modelling to the
strategy based on exact solutions. We argue that the efforts made and progress
achieved in this seemingly purely theoretical topic is of interest for phenomenology. 
\end{abstract}

\section{Introduction}

Nowadays one of the primary challenges to physics is to understand the phase structure of strong
interactions. An important goal of heavy-ion physics is thus to interpret the results of high-energy
collider experiments. It is a hard task, since one needs to follow the time-evolution
of the created matter in order to see the collective properties. The mean free path is small if 
temperature is high, (as first noted by Fermi~\cite{Fermi:1950jd}), so the idea arises naturally
to use hydrodynamics for this end. Hydrodynamics is almost the only way which dynamically
connects the initial conditions with the final state. As the first results from the RHIC
particle accelerator appeared, lots of models failed to describe the measurements. However,
many successful models were based on hydrodynamics, and this in-turn led to a firm
understanding that the created matter is an almost perfect liquid~\cite{Adcox:2004mh}. A typical feature of
the measured soft hadronic observables was the appearance of different scalings.
The strength of hydrodynamics lies in the fact that it relies only on the simple
assumption of local thermal equilibrium and local energy-momentum conservation,
and no physical scales are present, this leads to an easy explanation of scalings. 

\textit{The equations of relativistic hydrodynamics:}
In this subsection we briefly review the well-known equations of perfect fluid
relativistic hydrodynamics. The metric is $g^{\mu\nu}$, $u^{\mu}=\gamma(1,\v{v})$
is the four-velocity field, $\v{v}=v\v{n}$ is the three-velocity.
The pressure is denoted by  $p$\,, the energy density by $\varepsilon$\,, the temperature
by  $T$\,, and the entropy density by $\sigma$.
In high energy collisions, $\sigma$ is large compared to the net baryonic charge
density, so in the following we will not take conserved charges into account. The fundamental
equations are obtained by Landau's argumentation, which starts from the
conservation of energy-momentum and entropy density, expressed as
\bl{e:conservation}
\partial_\nu (\sigma u^\nu)  = 0 \quad,\quad \partial_\nu T^{\mu\nu} = 0 \quad,\quad 
T^{\mu\nu}=(\varepsilon+p)u^\mu u^\nu-pg^{\mu\nu} .
\ee
This form of the $T_{\mu\nu}$ energy-momentum tensor specifies the prefectness of the fluid.
The Euler equation and the energy equation follow as
\bea
(\varepsilon+p)u^{\nu}\partial_{\nu}u^{\mu}&=& \z{g^{\mu\rho}-u^{\mu}u^{\rho}}\partial_{\rho}p, \label{Reul} \\
(\varepsilon + p)\partial_{\nu}u^{\nu}+u^{\nu}\partial_{\nu}\varepsilon &=& 0. \label{RE}
\eea
These equations have to be supplemented by an equation of state (EoS), which connects $p$, $T$, and $\varepsilon$,
in order to have a closed set of equations. Assuming $\varepsilon=\kappa(T)p$, the $\kappa$ factor is $1/c_s^2$,
the inverse speed of sound. In most exact solutions one uses $\kappa=const.$

\textit{Exact vs. numerical solutions:}
Having the hydrodynamical equations, we can either solve them numerically or investigate them
analytically. They are nonlinear, and thus it is hard to find even particular analytic solutions.
So obviously, the main advantage of the numerical approach is that in this way one can in principle
use any type of initial conditions, and
calculate the corresponding final state observables. On the other hand, the similar advantage of
the analytic approach is also obvious: if one finds a suitable analytic solution, then one
can map not only a single initial condition but a manifold of them, and constrain
its parameters. Also, many classes of exact solutions are parametric solutions, naturally explaining scalings.
So our point is that analytic hydrodynamical solutions can also yield important insight into the dynamics. The interest
in this direction has somewhat revived in the last few years; we will first summarize the historical results,
then some recent developments.
 
\section{Historic results}

The most important and seminal two relativistic hydrodynamical solutions, the Landau-Khalatnikov solution
and the Hwa-Bjorken solution had great impact in the application of relativistic hydrodynamics to high-energy phenomena.

\textit{The Landau-Khalatnikov solution:}
The idea of relativistic hydrodynamics stems mostly from Landau. He also elaborated on Fermi's idea on the
applications~\cite{Landau:1953gs,Belenkij:1956cd}, and Kha\-lat\-ni\-kov gave the first analytic
solution to the relativistic hydrodynamical equations~\cite{Khalatnikov:1956xx}. 
This solution is an 1+1 dimensional, implicit, complicated one. We just highlight the main
notions and steps. What is needed, is the expression of the $T$ temperature and $\Omega$ fluid rapidity, defined
as $v=\tanh\Omega$, as a function of $t$ and $r$, the time and spatial coordinate, or of $x_+=t+r$ and $x_-=t-r$, the
lightcone coordinates. Rearranging the hydrodynamical equations a bit, one arrives at the conclusion that
the key to the solution is a potential, $\Phi(x_+,x_-)$, with $\partial_+\Phi=Te^\Omega$, $\partial_-\Phi=Te^{-\Omega}$,
and $\Phi$ can be calculated from its Legendre-transform $\chi\z{Te^\Omega,Te^{-\Omega}}=\Phi-x_+Te^\Omega-x_-Te^{-\Omega}$,
which satisfies the linear Khalatnikov-equation: 
\be
\partial_\theta^2\chi\z{\theta,\Omega}+\z{\kappa-1}\partial_\theta\chi\z{\theta,\Omega}-\kappa
\partial_\Omega^2\chi\z{\theta,\Omega}=0 ,
\ee
where $\theta=\ln T$ was used. Now the solution of this equation can be written up with integral-formulas
using the Green-function formalism (see e.g.~\cite{Beuf:2008vd}); the essence of the Landau-Kha\-lat\-ni\-kov-solution is the fully stopped finite
piece of matter initial condition. It yields approximately Gaussian rapidity distribution for the produced particles, which
is a realistic prediction.

\textit{The Hwa-Bjorken solution:}
Contrasted to the Landau-Khalatnikov solution, the Hwa-Bjorken solution (originally formulated by Hwa~\cite{Hwa:1974gn}, discussed by
many others, rediscovered and fully exploited by Bjorken~\cite{Bjorken:1982qr}) provides an over-simplified picture of the 1+1 dimensional
dynamics. It uses the $\tau$ and $\eta$ Rindler-coordinates: the time $t$ and spatial coordinate $r$ is expressed as
$t=\tau\cosh\eta$, $r=\tau\sinh\eta$. The core assumption (valid at infinite collision energies) is the boost-invariance,
i.e.\@ that $\sigma$ and $T$ are independent of $\eta$, and indeed, the simple
\be
v=r/t \quad , \quad \sigma_0/\sigma=\tau_0/\tau 
\ee
forms give an accelerationless solution of the hydrodynamical equations. The expression of the temperature depends on
the actual value of $\kappa$. This solution leads to a flat rapidity distribution, thus although
it can be used approximately to various estimates, it needs a correction.

\section{Recent results}

  \mbox{}\textit{Nonrelativistic models:} Although relativistic effects are more than essential in high-energy experiments, the nonrelativistic case also
deserves a brief summary here: the equations are much simper, and allow for more
exact solutions~\cite{Csorgo:1998yk,Akkelin:2000ex,Csorgo:2001ru,Csizmadia:1998ef,Csorgo:2001xm}. A pretty general family is described
in Ref.~\cite{Csorgo:2001ru}, with a self-similar ellipsoidal velocity and temperature profile. It contains some of the other solutions
as special cases. It serves as a base of the Buda-Lund model, which is successful in describing particle spectra, correlations, and
their scalings~\cite{Csanad:2003qa,Csorgo:1995bi}.

It is also worthwhile to mention this exact solution because (as far as we know) this is the only one which can be generalized for arbitrary
temperature-dependent speed of sound: if one assumes Gaussian density profile and spatially constant temperature, then one gets a parametric
solution for any $\kappa(T)$ function. This result --- though non-relativistic --- is unique, and makes possible to use any QCD-inspired EoS.
One would naturally look for similar relativistic solutions.

\textit{Relativistic accelerationless solutions:}
The generalization of the Hwa-Bjorken solution to arbitrary number of spatial dimensions seems a straightforward direction of development,
although it was a formidable task~\cite{Csorgo:2003rt,Csorgo:2003ry}. These solutions are also the relativistic equivalents of
the nonrelativistic solutions mentioned in the previous subsection. They have an accelerationless,
spherically symmetric velocity profile: $v=\v{r}/t$. The pressure is $p=nT$, with some conserved charge $n$. Ellipsoidal
profiles are allowed in the forms of $n$ and $T$ as
\be
n=n_0\z{\frac{\tau_0}{\tau}}^{3}\rec{\c{T}(S)} \quad,\quad 
T=T_0\z{\frac{\tau_0}{\tau}}^{3\rec{\kappa}}\c{T}(S)\quad,\quad S=\rec{t^2}\z{\frac{x^2}{A^2}+\frac{y^2}{B^2}+\frac{z^2}{C^2}} , 
\ee 
where $\c{T}$ is an arbitrary function of the ellipsoidal scaling variable $S$, with principal axes $A$, $B$, and $C$ 
in the directions $x$, $y$, and $z$. In Ref.~\cite{Csorgo:2003ry} other generalizations are also found, e.g. to
hyperbolic profiles, and Ref.~\cite{Sinyukov:2005mn} shows a slight generalization, where
even the velocity field can show more general, ellipsoidal symmetry, but still without any acceleration.
Other important accelerationless solutions were presented in Refs.~\cite{Biro:1999eh,Biro:2000nj}.

\textit{Accelerating solutions:} There were no known examples of exact explicit and accelerating solutions until
recently an interesting class of spherically symmetric solutions emerged~\cite{Csorgo:2006ax,Nagy:2007xn}: as
a generalization of the Hwa-Bjorken solution, one finds that the
\bl{e:solution}
v=\tanh\z{\lambda\eta} \quad,\quad p=p_0\z{\tau_0/\tau}^{\lambda d(\kappa+1)/\kappa}\cosh^{-(d-1)\Phi_\lambda}\z{\eta/2}
\ee  
expressions are indeed solutions of the hydrodynamical equations, for certain values of the real parameters $\lambda$, $d$,
$\Phi_\lambda$ and $\kappa$: $\kappa$ is the (constant) inverse speed of sound, $d$ is the number of spatial dimensions, $\lambda$ is a
parameter of the solutionr; for $\lambda=1$ the Hwa-Bjorken solution is recovered, for $\lambda\neq 1$ the solution is accelerating.
The $\Phi_\lambda$ parameter is introduced because for some choices of $\lambda$ the pressure depends on $\eta$ as well.
The allowed parameter sets are listed in Table~\ref{t:solution}. They have many interesting properties; for a detailed explanation, see
Ref.~\cite{Nagy:2007xn}. The $\kappa=1$, $d=1$ solutions, on the other hand, allow to an easy approximate calculation of the rapidity
distribution~\cite{Nagy:2007xn}. These distributions qualitatively agree with the observed peaked (Gaussian) structure, and the
$\lambda$ parameter can be extracted from a fit to measured data with acceptable statistical significance, thus these solutions
serve as a means to improve Bjorken's original estimate~\cite{Bjorken:1982qr} of the initial energy density: the work done by
the fluid (because of acceleration) and the shift in the estimated origin of the trajectories caused by the presence of
acceleration leads to the conclusion that the Bjorken
estimate needs to be corrected by a factor greater than $1$: for $\sqrt{s_{NN}}=200$GeV Au+Au collisions, from rapidity distributions
measured by the BRAHMS collaboration, one gets not less than a factor of $2.0\pm 0.1$ correction, and taking the softness of the EoS
into account, a conjectured correction factor of $2.9\pm 0.2$~\cite{Csorgo:2006ax}. This result is important in the interpretation
of experimental data in terms of advanced estimates of the initial energy densities. A somewhat less important estimate can also
be made more precise: the life-time of the reaction increases by about 20\% with taking the acceleration into account in this way~\cite{Csorgo:2006ax}.
\begin{table}
\begin{center}
\begin{tabular}{|c|c|c|c|c|}
  \hline
  Case & $\lambda$ &       $d$             & $\kappa$ &         $\phi_{\lambda}$            \\ \hline
  (a)  & $2$             & $\in\mathbb{R}$ & $d$              & $0$                         \\
  (b)  & $\rec{2}$       & $\in\mathbb{R}$ & $1$              & $\frac{\kappa + 1}{\kappa}$ \\
  (c)  & $\frac{3}{2}$   & $\in\mathbb{R}$ & $\frac{4d-1}{3}$ & $\frac{\kappa + 1}{\kappa}$ \\
  (d)  & $1$             & $\in\mathbb{R}$ & $\in\mathbb{R}$  & $0$                         \\
  (e)  & $\in\mathbb{R}$ & $1$             & $1$              & $0$                         \\  \hline
\end{tabular}
\caption{Allowed parameters for the family of accelerating solutions of Eq.~\r{e:solution}.}\label{t:solution}
\end{center}
\end{table}
It should be noted that in the case of stiff EoS, $\kappa=1$, not only these solutions with the $\lambda$ parameter,
but the general explicit solution can be obtained~\cite{Nagy:2007xn} because of an analogy to a linear wave-equation.
For multi-dimensional flows, this idea resulted in a broad class of new general solutions~\cite{Borshch:2007uf}, although only for this particular
EoS, and it is not clear how these results could be generalized for any other.

\textit{Harmonic flows in $1+1$ dimensions:}
Another recent approach toward new solutions was a generalization of the Bjorken ansatz (the boost invariance)
to a harmonic ansatz: $\partial_+\partial_-\Omega=0$. A new class of solutions
is obtained when substituted into the hydrodynamical equations~\cite{Bialas:2007iu}:
\be
p=p_0\exp\kz{-\frac{\z{1+\kappa}^2}{4\kappa}\z{l_+^2+l_-^2}+\frac{\kappa^2-1}{2\kappa}l_+l_-} \quad,\quad \Omega=\rec{2}\z{l_+^2-l_-^2} .
\ee
The notations are 
\be
l_\pm(x_\pm)=\sqrt{\ln F_\pm} \quad,\quad z_\pm=h\int^{F_\pm}\frac{dx}{\sqrt{\ln x}} ,
\ee
where $h$ is an arbitrary constant. This solution, although not fully explicit, is very interesting, since it
interpolates between the Landau and the Bjorken pictures (fixed $h$, $l_\pm\to\infty$, and $h\to 0$, respectively).
If one calculates the entropy density per unit rapidity, which is proportional to the observable particle distribution, it depends
on the assumed freeze-out surface, but in general it is approximately Gaussian~\cite{Bialas:2007iu}. More general expressions for the
entropy flow $\td{S}{n}$, based partially on the Khalatnikov method, were discussed recently in Ref.~\cite{Beuf:2008vd}.

\section{Summary: where we are now and where to go}

The interest in the numerical simulations of relativistic hydrodynamics is steadily growing: it seems obvious to almost everyone that
hydrodynamics is the correct tool to describe high-energy collective phenomena. We see now that a similar common interest begins
to arise towards exact solutions. With simple examples we tried to demonstrate that there are many new and interesting solutions, and
that these are of phenomenological importance: if one has a solution with a few adjustable fit parameters, it gives invaluable
insight into the dynamics and yields advanced estimates for the initial conditions (such as energy density, life-time).

Finding exact solutions to the hydrodynamical equations, however, is a difficult problem, and needs lots of effort. For instance, there is
no known solution in more than one spatial dimensions with a little bit general equation of state. Similarly, no accelerating solutions
are known  which go beyond spherical symmetry. The quest for such solutions (e.g. for an ellipsoidally symmetric one) might lead to a
more accurate description of the observables, and thus test the perfectness of the fluid and the used equation of state.

The support of OTKA T49466 and NK73143 grants is gratefully acknowledged. We thank to the organizers of ISMD 2008
for their kind hospitality and support.
 
\begin{footnotesize}
\bibliographystyle{ismd08} 
{\raggedright
\bibliography{ismd08_marton_nagy}
}
\end{footnotesize}
\end{document}